\documentclass[%
reprint,
superscriptaddress,
amsmath,amssymb,
aps,
prl,
]{revtex4-2}

\usepackage{graphicx}
\usepackage{dcolumn}
\usepackage{bm}
\usepackage[mathlines]{lineno}
\usepackage{braket}
\usepackage[dvipsnames]{xcolor}
\usepackage{amsmath,amssymb,amsfonts} 
\usepackage[utf8]{inputenc}
\usepackage{appendix}
\usepackage{mhchem}
\usepackage{makecell}
\usepackage{float}
\usepackage{bbm}
\usepackage{nicefrac}
\usepackage{wasysym}
\usepackage[normalem]{ulem}

\usepackage{hyperref}
\hypersetup{
	colorlinks=true,
	linkcolor=blue,
	filecolor=blue,      
	urlcolor=blue,
	citecolor=blue,
}
\def\Red#1{\textcolor{red}{#1}}
\def\Blue#1{\textcolor{blue}{#1}}

\begin{document}
\title{Thickness-Independent Quantum Geometric Responses Driven by \\ Interlayer Antiferroic Coupling}

\author{Zhiming Xu}
\affiliation{School of Physics, Peking University, Beijing 100871, China}

\author{Huaqing Huang}
\email[contact author: ]{huaqing.huang@pku.edu.cn}
\affiliation{School of Physics, Peking University, Beijing 100871, China}
\affiliation{Collaborative Innovation Center of Quantum Matter, Beijing 100871, China}
\affiliation{Center for High Energy Physics, Peking University, Beijing 100871, China}


\begin{abstract}
Two-dimensional ferroic materials exhibit rich and intriguing physical phenomena, but their response properties generally depend sensitively on thickness, requiring precise layer-number control and thereby limiting practical applications. Here, we propose a general strategy for realizing thickness-independent quantum geometric responses through symmetry engineering induced by interlayer antiferroic coupling. Using spatial-dependent symmetry analysis, we show that thickness-independent behavior emerges when the symmetry breaking required for a given response is generated by interlayer antiferromagnetic (AFM) or antiferroelectric (AFE) coupling, without invoking topological mechanisms. Our first-principles calculations predict that multilayer MnS in the G-type AFM configuration exhibits a surface-dominated anomalous Hall effect, whose thickness-independent behavior can be significantly influenced by the stacking order. We further propose design principles for achieving thickness-independent anomalous and nonlinear Hall effects driven by interlayer AFE coupling, and suggest potential applications in distinguishing magnetic structures. Our findings open a new route towards robust functional devices based on antiferroic materials.
\end{abstract}

\maketitle

\textit{Introduction.}---Two-dimensional (2D) ferroic materials have attracted extensive research efforts because of their abundant physical properties and promising applications in next-generation technological devices \cite{gong2019two,huang2020emergent,zhang2023ferroelectric,wang2023towards,hu2019progress,man2023ferroic}. They involve one or multiple ferroic orders, such as magnetism and (anti)ferroelectricity, down to the atomic-scale thickness. Due to the unique broken symmetries, a variety of fascinating physical phenomena have been discovered in 2D ferroic materials, including various types of Hall effects \cite{deng2018gate,deng2020quantum,jiang2020concurrence,ma2019observation,kang2019nonlinear,wang2023quantum,gao2023quantum,yu2025quantummetric}, nonlinear optics \cite{autere2018nonlinear,sun2019giant,xu2020spontaneous}, magnetoelectric \cite{gao2024giant,aoki2024giant,zhang2025magnetoelectric} and magneto-optical phenomena \cite{gong2017discovery,huang2017layer,ju2021possible}. Therefore, 2D ferroic materials provide a promising platform for exploring emergent quantum phenomena and developing novel functional devices.

In general, however, the response properties of 2D systems are strongly dependent on the layer thickness owing to the quantum confinement effect \cite{pertsev2007thickness,li2017layer,deng2022layer}. As a result, achieving targeted device performance often requires atomic-scale, precise control of the layer number, which poses a major challenge for practical implementation. 
In contrast, thickness-independent responses would greatly ease device fabrication, improve performance robustness, and broaden application scope.
A notable example is provided by topological materials \cite{bansal2012thickness, hasan2010colloquium, qi2011topological, chang2023colloquium}, in which transport properties are tied to topological invariants and are dominated by surface states. 
Yet such thickness-independent responses are typically confined to a narrow topological gap, which severely restricts their persistence against finite temperature, doping, and external perturbations
\cite{pan2020probing, liu2023magnetic}. Very recently, surface-dominated nonlinear responses have also been reported in antiferromagnetic (AFM) materials \cite{zhou2024skin,das2025surface}, where inner-layer contributions are nearly suppressed by locally preserved symmetry, leading to thickness-independent behavior. These findings suggest that interlayer antiferroic interactions may provide an alternative route to engineering unusual spatially dependent responses.

In this work, we propose a general strategy to realize thickness-independent quantum geometric responses through symmetry engineering enabled by interlayer antiferroic coupling. Considering interlayer interactions are short-ranged, we employ spatial-dependent symmetry analysis to demonstrate that the thickness-independent behavior can arise, without relying on topological effects, when the symmetry prohibiting the response is broken by interlayer AFM or antiferroelectric (AFE) coupling. Taking the layered AFM material MnS as a representative, we reveal that interlayer AFM configuration-induced symmetry breaking leads to surface-dominated anomalous Hall effect (AHE), and the resulting thickness-independent behavior can be effectively controlled by the stacking order. We further formulate design principles for thickness-independent anomalous and nonlinear Hall effects driven by interlayer AFE coupling and discuss potential applications of such responses. Our work sheds light on designing practical functional devices based on AFM spintronics and AFE electronics.

\textit{Material design principle.}---We begin with a general analysis of thickness-dependent behaviors of response properties. The Hamiltonian of an $N$-layer system can be decomposed into intralayer and interlayer parts as
\begin{equation}
H=\sum_{i=1}^N H^{\mathrm{intra}}_{i}(\boldsymbol{m}_i)+\sum_{i < j} H^{\mathrm{inter}}_{i,j}(\boldsymbol{m}_i, \boldsymbol{m}_j),
\end{equation}
where $i$ and $j$ are the layer indices, $\boldsymbol{m}_i$ is the ferroic order parameter of the $i$-th layer, $H^{\mathrm{intra}}_{i} = H_{i,i}(\boldsymbol{m}_i, \boldsymbol{m}_i)$, $H^{\mathrm{inter}}_{i,j} = H_{i,j}(\boldsymbol{m}_i, \boldsymbol{m}_j)+H_{j,i}(\boldsymbol{m}_j, \boldsymbol{m}_i)$ with $i \neq j$. Similarly, the quantum geometric response coefficient $\sigma$, which can be related to quantum geometric quantities such as Berry curvature, quantum metric, and their derivatives \cite{yu2025quantum,liu2025quantum}, can be decomposed into layer-resolved contributions
\begin{equation}
\sigma=\sum_{i=1}^N \sigma^{\mathrm{intra}}_{i}(\boldsymbol{m}_i)+\sum_{i \neq j} \sigma^{\mathrm{inter}}_{i,j}(\boldsymbol{m}_i, \boldsymbol{m}_j),
\end{equation}
where $\sigma^{\mathrm{intra}}_{i}$ ($\sigma^{\mathrm{inter}}_{i,j}$) denotes the contribution arising from intra-layer ($i \to i$) and inter-layer ($i \to j$) interaction, respectively. These quantities are obtained by projecting the velocity operator onto individual layers [see Supplemental Material (SM) for details \cite{SM}].
Considering interlayer interactions across van der Waals gaps are short-ranged, we adopt the nearest-neighbor layer approximation, assuming only the intralayer and nearest-neighbor interlayer interactions make the dominant contributions
\begin{equation}
\begin{split}
H &\approx \sum_{i=1}^N H^{\mathrm{intra}}_{i}(\boldsymbol{m}_i)+\sum_{i=1}^{N-1} H^{\mathrm{inter}}_{i,i+1}(\boldsymbol{m}_i, \boldsymbol{m}_{i+1}), \\
\sigma &\approx \sum_{i=1}^N \sigma^{\mathrm{intra}}_{i}(\boldsymbol{m}_i)+2\sum_{i=1}^{N-1} \sigma^{\mathrm{inter}}_{i,i+1}(\boldsymbol{m}_i, \boldsymbol{m}_{i+1}).
\end{split}
\end{equation}

\begin{figure}[tb]
\includegraphics[width=\linewidth]{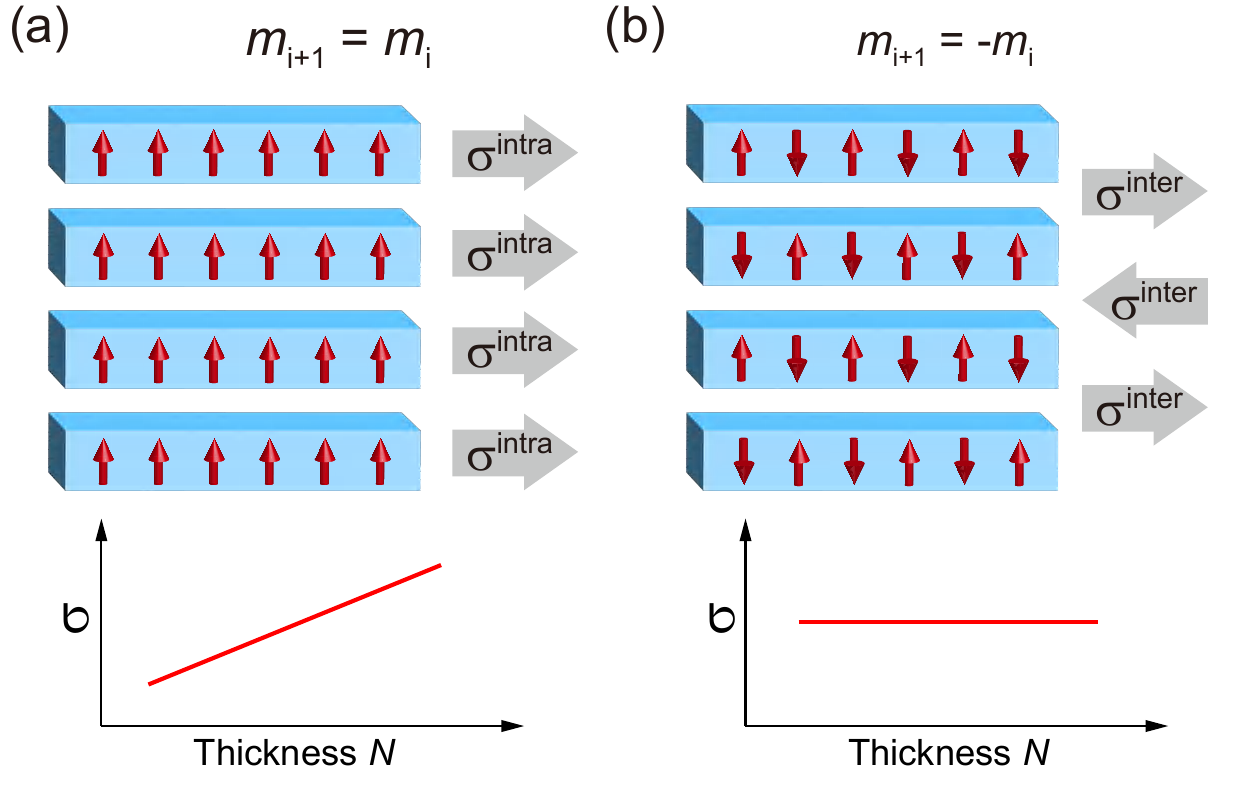}
\caption{\label{Fig1}Two typical thickness-dependent response behaviors in the weak interlayer coupling limit. (a) For materials with identical ferroic order parameters in each layer, the response signal approximately linearly scales with the thickness. (b) For materials with interlayer antiferroic configuration, the response signal is nearly independent of the thickness.}
\end{figure}

For systems with a uniform ferroic order parameter across layers---whether ferroic or nonferroic---each layer contributes nearly equally to the intralayer response. 
In the weak interlayer coupling limit, where the multilayer can be treated as multiple separated single layers with negligible interlayer interactions, the total response signal approximately equals the sum of the intralayer contribution of each layer, rendering $\sigma$ an extensive quantity that scales linearly with thickness [Fig.~\ref{Fig1}(a)]. 

The situation is qualitatively different for interlayer antiferroic order, $\boldsymbol{m}_{i+1}=-\boldsymbol{m}_i$. In this case, the symmetry of the system depends on layer parity because the magnetic or electric dipoles alternate from layer to layer. 
A symmetry $\mathcal{S}$ may be preserved in odd-layer systems and the bulk, yet broken in even-layer systems. For a quantum geometric response that is $\mathcal{S}$-odd ($\hat{\mathcal{S}}\sigma = -\sigma$), the short-ranged nature of interlayer coupling implies that $\mathcal{S}$ is approximately preserved in interior layers, suppressing their intralayer contribution $\sigma^{\mathrm{intra}}_{i}$ and enforces a cancellation of interlayer contributions: $\sigma^{\mathrm{inter}}_{i,i+n} = - \sigma^{\mathrm{inter}}_{i,i-n}$. With the nearest-neighbor layer approximation, interior layers contribute negligibly, and surface layers dominate the response: $\sigma \approx \sigma^{\mathrm{inter}}_{1,2}(\boldsymbol{m}_1,-\boldsymbol{m}_1) + \sigma^{\mathrm{inter}}_{N-1,N}(-\boldsymbol{m}_N, \boldsymbol{m}_N)$. For compensated antiferroic systems with even layer number, $\boldsymbol{m}_N = -\boldsymbol{m}_1$, yielding $\sigma \approx 2\sigma^{\mathrm{inter}}_{1,2}$---a quantity independent of layer thickness. The response thereby behaves as an intensive property of the system [Fig.~\ref{Fig1}(b)].

The above analysis yields a concise set of conditions for thickness-independent quantum geometric responses: 
(i) The interlayer coupling is antiferroic and short-ranged. (ii) The response is symmetrically forbidden for the monolayer and bulk, while allowed for the multilayer. 
This mechanism is topologically trivial---it does not rely on any gap topology---in contrast to surface states in topological insulators. If the interlayer coupling is strong enough to induce a topological phase transition, additional topology-driven contributions may arise, which are beyond the scope of this work. 

\begin{table*}[t]
\centering
\caption{Symmetry requirements and material candidates of different types of thickness-independent response properties.}\label{Tab1}
\setlength{\tabcolsep}{9pt}{
\renewcommand\arraystretch{1.9}
\begin{tabular}{c|c|c|c}
\hline\hline
Response & Preserved symmetries $\mathcal{S}$ & Broken symmetries $\mathcal{S}^{\prime}$ & Material candidates \\ \hline
Anomalous Hall effect & $\mathcal{PT}$, $M_z\mathcal{T}$, $C_{2v}$, $C_{nz}\mathcal{PT}$ & $\mathcal{T}\tau_v$, $M_v$, $C_{nz}\mathcal{T}$ & \makecell{G-type AFM [Fig.~\ref{Fig2}(a)], \\ A-type AFM [Fig.~S1(b) \cite{SM}]} \\ \hline
$\mathcal{T}$-even second-order effects & $\mathcal{P}$, $\mathcal{PT}$ & $C_{nz}$, $C_{nz}\mathcal{T}$ & G-type AFE [Fig.~\ref{Fig4}(b)] \\ \hline
$\mathcal{T}$-odd second-order effects & $\mathcal{P}$, $M_z\mathcal{T}$ & $\mathcal{T}\tau_v$, $C_{nz}$, $C_{3/4/6z}\mathcal{T}$ & \makecell{A-type AFM: CrI$_3$ (Ref. \cite{zhou2024skin}) \\ and CrSBr (Ref. \cite{das2025surface}) }  \\ \hline\hline
\end{tabular}}
\end{table*}

We now derive the symmetry requirements for monolayers that enable thickness-independent quantum geometric responses. For clarity, we consider AA stacking to eliminate any additional symmetry breaking that may arise from the stacking configuration. Interlayer antiferroic coupling can break magnetic layer-group symmetries that involve atomic exchange between layers (see detailed derivations in SM \cite{SM}). The relevant symmetries can be expressed as
\begin{equation}
\{\mathcal{P}, M_z, C_{2v}, C_{nz}\mathcal{P}\} \circ \{I,\mathcal{T}\},
\end{equation}
where $\mathcal{P}$ ($\mathcal{T}$) denotes inversion (time-reversal) symmetry, $M_z$ is mirror symmetry with respect to the $z$ direction, $C_{2v}$ ($C_{nz}$) represents twofold ($n$-fold) rotation about the in-plane (out-of-plane) axis, and $I$ is the identity operation. 
Among these symmetries, we identify those that forbid the response in the monolayer and denote them as $\mathcal{S}$, as summarized in Table~\ref{Tab1}. To ensure a finite response in the compensated antiferroic multilayer, all other symmetries that prohibit the response but cannot be broken by interlayer antiferroic coupling must already be broken in the monolayer, which we denote as $\mathcal{S}'$. Consequently, the symmetry requirement for thickness-independent responses is that the monolayer preserves at least one $\mathcal{S}$ symmetry, which is then broken by interlayer antiferroic coupling in multilayers, while simultaneously breaking all $\mathcal{S}'$ symmetries (Table~\ref{Tab1}).

\textit{Thickness-independent AHE driven by interlayer AFM coupling.}---
We take the AHE as a representative to explore potential material candidates. As listed in Table \ref{Tab1}, thickness-independent AHE can be expected if the monolayer preserves one of the symmetries $\{\mathcal{PT}$, $M_z\mathcal{T}$, $C_{2v}$, $C_{nz}\mathcal{PT}\}$, meanwhile breaks $\mathcal{T}\tau_v$ ($\tau_v$ is in-plane translation), $M_v$ (mirror symmetry along the in-plane direction) and $C_{nz}\mathcal{T}$. Based on these symmetry requirements, we propose two prototype material classes capable of hosting thickness-independent AHE (see Fig.~S1 in SM \cite{SM}): (1) G-type AFM material whose monolayer is a bipartite AFM with $\mathcal{PT}$ symmetry. (2) In-plane magnetized H-phase transition metal dichalcogenide with A-type AFM order, whose monolayer preserves $M_z\mathcal{T}$ and $C_{2v}$ symmetries.

\begin{figure}[b]
\includegraphics[width=\linewidth]{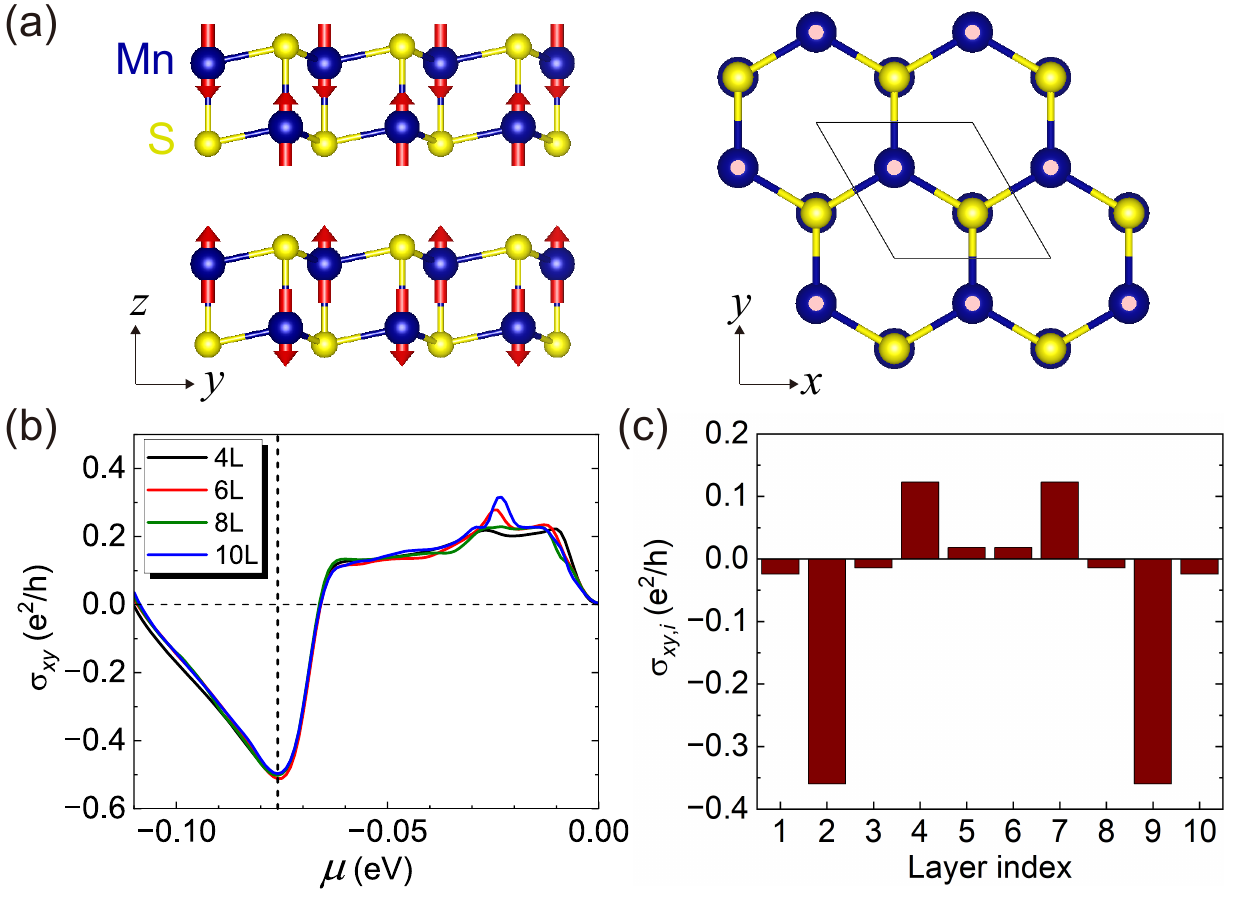}
\caption{\label{Fig2}(a) Side and top views of out-of-plane magnetized MnS with G-type AFM configuration. Red arrows indicate magnetic moments. (b) Anomalous Hall conductivity $\sigma_{xy}$ as a function of chemical potential $\mu$ with the thicknesses of 4, 6, 8, and 10 layers. The zero point of $\mu$ is set at the valence band maximum. The unit of $\sigma_{xy}$ is quantum conductivity $e^2/h$. (c) Layer-resolved $\sigma_{xy}$ of a 10-layer MnS at $\mu = -0.076$ eV, which corresponds to the vertical dashed line in (b).}
\end{figure}

To validate the reliability of the above analyses, we perform first-principles calculations on the layered antiferromagnet MnS, which has recently attracted research interest owing to its robust AFM ordering and novel Hall responses \cite{shahid2022monolayer,aapro2021synthesis,wang2023intrinsic,long2025anomalous}. As shown in Fig.~\ref{Fig2}(a), MnS crystallizes in a layered hexagonal structure, whose monolayer consists of an AA-stacked buckled honeycomb bilayer formed by two Mn-S sublayers with inverted sublattices. With interlayer AFM configuration-induced $\mathcal{PT}$ symmetry breaking, multilayer MnS with G-type AFM order satisfies the symmetry requirements presented above. 
Our first-principles results show that bulk MnS is an insulator with a band gap of 1.7 eV, and less dispersive bands along the $\Gamma$-Z line indicate that the interlayer coupling is weak (see Fig.~S2 in SM \cite{SM}). Then we investigate the evolution of anomalous Hall conductivity $\sigma_{xy}$ of out-of-plane magnetized MnS with respect to the layer thickness. As shown in Fig.~\ref{Fig2}(b), $\sigma_{xy}$ remains nearly unchanged when the thickness increases from four to ten layers for chemical potential $\mu$ between -0.11 eV and 0 eV, demonstrating a clear thickness-independent response. 

To present the spatial dependent contribution of this unusual response, we further calculate the layer-resolved anomalous Hall conductivity $\sigma_{xy,i}$, which is obtained by collecting all the response coefficients involving the interaction from the $i$-th layer (see details in SM \cite{SM}). Figure \ref{Fig2}(c) shows the calculated $\sigma_{xy,i}$ of a 10-layer MnS at the negative conductivity peak, which corresponds to $\mu$ = -0.076 eV. $\sigma_{xy,i} = \sigma_{xy,11-i}$ is due to the $\mathcal{P}$ symmetry in the multilayer. The outermost two layers ($i$ = 1,2 or 9,10) contribute dominantly, indicating the thickness-independent response is surface-dominated and exhibits a skin-effect-like spatial profile \cite{zhou2024skin}, in agreement with our symmetry analysis. We also examine MnS with magnetization along the $y$ axis (see Fig.~S3 \cite{SM}) and find that, although the magnitude of the response is much smaller, the AHE remains surface-dominated and nearly independent of thickness, further supporting the generality of the mechanism.

\begin{figure*}[htb]
\includegraphics[width=\linewidth]{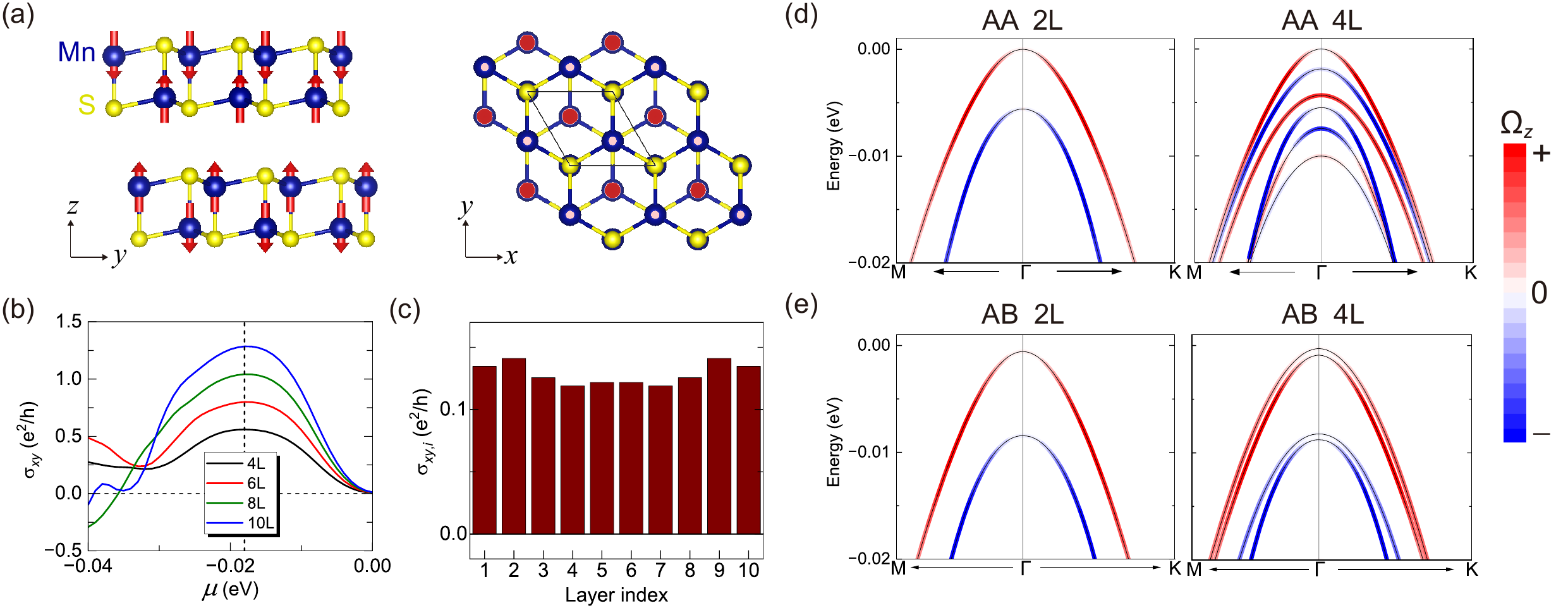}
\caption{\label{Fig3}(a) Side and top views of AB-stacked MnS. Red arrows indicate magnetic moments. (b) $\sigma_{xy}$ as a function of $\mu$ with the thicknesses of 4, 6, 8, and 10 layers. (c) Layer-resolved $\sigma_{xy}$ of a 10-layer AB-stacked MnS at $\mu$ = -0.018 eV, which corresponds to the vertical dashed line in (b). (d) Band-resolved Berry curvature of 2-layer and 4-layer MnS with AA stacking. (e) Band-resolved Berry curvature of 2-layer and 4-layer MnS with AB stacking.}
\end{figure*}

\textit{Stacking-controlled thickness-dependent response behaviors.}---Due to weak van der Waals interlayer interactions, layered materials possess a versatile stacking degree of freedom, which has a significant influence on their physical properties \cite{wu2021sliding,ji2023general,pan2024general,he2025switching}. We therefore examine how stacking affects the thickness dependence of the AHE by considering AB-stacked MnS.
As shown in Fig.~\ref{Fig3}(a), for the AB stacking, atomic and magnetic structures of the (2$i$)-th layer are constructed by performing the $\hat{\mathcal{T}}\hat{\tau}_{[1/3, -1/3, 0]}\hat{\tau}_z$ operation to the (2$i-$1)-th layer, where $\hat{\tau}_{[1/3, -1/3, 0]}$ is in-plane fractional translation operator along the long diagonal of the unit cell and $\hat{\tau}_z$ is out-of-plane translation operator. Such stacking order breaks the $\mathcal{PT}$ symmetry, allowing the AHE in both bulk and multilayer and violating our symmetry condition required for thickness-independent response. Consistent with this expectation, Figs. \ref{Fig3}(b) and \ref{Fig3}(c) show that $\sigma_{xy}$ scales approximately linearly with thickness when the system is doped near the valence band maximum (VBM), with nearly equal contributions from each layer.

The dramatic change in thickness dependence induced by interlayer relative displacement can be understood from the electronic structure. 
To elucidate the underlying microscopic origin, we calculate partial change densities and band-resolved Berry curvatures for two-layer (2L) and four-layer (4L) MnS under both stackings. 
Partial charge density analysis demonstrates that for both stackings, the electronic states near the VBM are contributed by the orbitals across the van der Waals gap (see Figs. S4 and S5 in SM \cite{SM}), reflecting interlayer hybridization and contributing to $\sigma^{\mathrm{inter}}_{i,i+1}$. For the AA stacking, additional subbands generated by increasing thickness carry Berry curvatures of opposite sign, as shown in Fig.~\ref{Fig3}(d). Since $\sigma_{xy}$ is obtained by summing Berry curvature over all occupied states, these additional contributions largely cancel, yielding a thickness-independent AHE consistent with the symmetry analysis.  
In contrast, AB stacking produces two symmetrically inequivalent bilayers: A-B and B-A. For the 4L A-B-A-B sequence, the electronic states near the VBM originate primarily from the A-B bilayers (see Fig.~S5 \cite{SM}). The two A-B bilayers in 4L can be related by the $\mathcal{P}$ symmetry and therefore host subbands with the same sign of Berry curvature, as shown in Fig.~\ref{Fig3}(e), leading to an approximately linear increase of $\sigma_{xy}$ with thickness. 

We further investigate the thickness-dependent response property of MnS with other stacking orders, including ABC, AC, AA$^{\prime}$, AB$^{\prime}$, and AC$^{\prime}$ stackings (see Fig.~S6 \cite{SM}). Our calculations show that the stacking order can effectively modulate both the magnitude and thickness dependence of the AHE in MnS, all of which can be consistently explained by our proposed symmetry condition. These results highlight stacking as a powerful approach for controlling thickness-dependent response properties in layered materials and provide new opportunities for designing slidetronics-based functional devices.

\textit{Thickness-independent responses driven by interlayer AFE coupling.}---The symmetry condition developed above can be generalized using the electric-magnetic duality between antiferromagnetic (AFM) and antiferroelectric (AFE) orders. Interlayer AFE coupling, therefore, provides an alternative route to realize thickness-independent quantum geometric responses. 

\begin{figure}[tb]
\includegraphics[width=\linewidth]{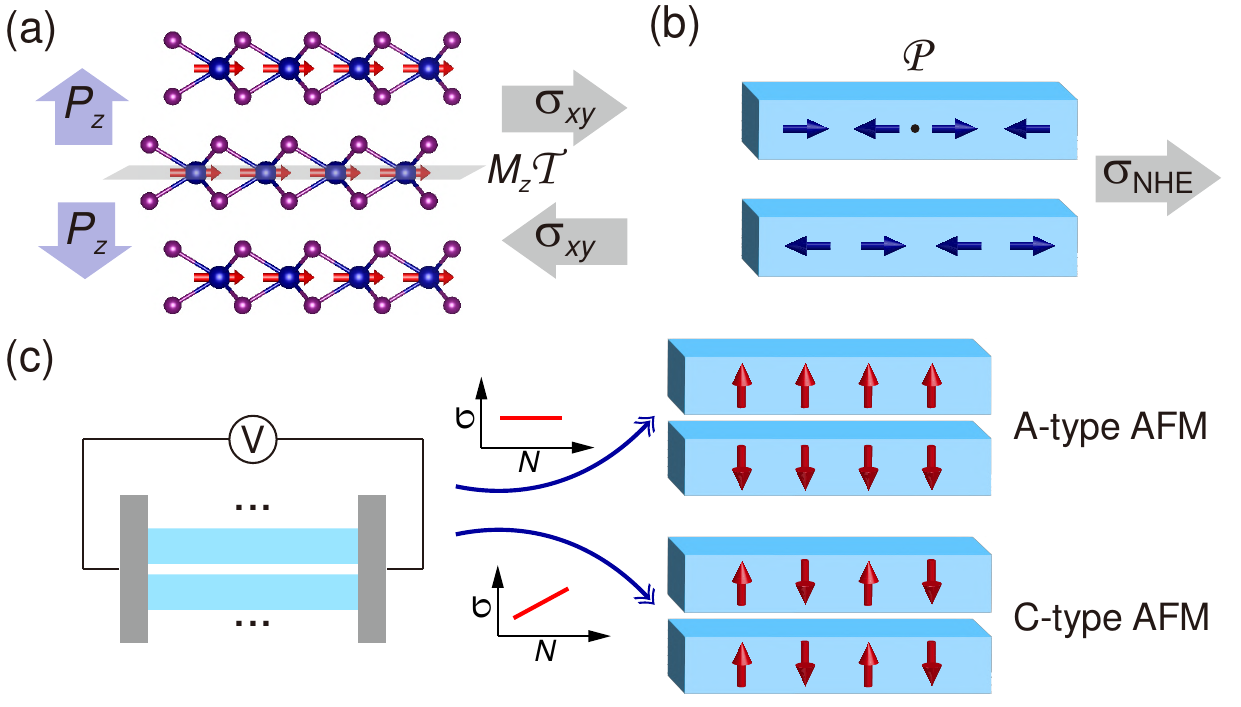}
\caption{\label{Fig4}(a) Atomic and magnetic structures of in-plane magnetized AB-stacked transition metal dichalcogenide. Red arrows indicate magnetic moments. (b) Schematic diagram of the G-type AFE material. Blue arrows indicate local electric dipoles. (c) Schematic diagram of using the thickness-dependent response behavior to distinguish different AFM configurations.}
\end{figure}

First, considering AHE driven by interlayer AFE coupling. As discussed above, thickness-independent AHE can be achieved by interlayer AFM coupling-induced $M_z\mathcal{T}$ symmetry breaking. Here, we utilize interlayer stacking configurations instead of magnetic configurations. As shown in Fig.~\ref{Fig4}(a), finite out-of-plane electric polarization $P_z$ and $\sigma_{xy}$ are allowed in AB-stacked transition metal dichalcogenide bilayer due to the broken $M_z\mathcal{T}$ symmetry. Notably, A-B and B-A bilayers are related by the $\hat{M}_z\hat{\mathcal{T}}$ operator, which can reverse both $P_z$ and $\sigma_{xy}$. Then, thickness-independent AHE driven by interlayer AFE configuration can be realized in the AB-stacked multilayer. 

Second, considering $\mathcal{T}$-even nonlinear Hall effect (NHE) in G-type AFE materials. $\mathcal{T}$-even NHE induced by Berry curvature dipole can arise in $\mathcal{T}$-invariant $\mathcal{P}$-broken materials \cite{sodemann2015quantum}. As shown in Fig.~\ref{Fig4}(b), for G-type AFE materials with both intralayer and interlayer AFE configurations, the $\mathcal{P}$ symmetry is preserved in monolayer and bulk, while broken in even layers. Then the NHE is forbidden (allowed) in monolayer and bulk (even layers), following our proposed symmetry condition. Thus, thickness-independent NHE driven by interlayer AFE coupling can be expected.

\textit{Potential applications.}---Thickness-independent response properties hold promising device application prospects. First, the response signals remain stable down to the few-layer limit, beneficial for device miniaturization. Second, because the response does not depend on the precise layer number, fabrication tolerances are significantly relaxed, improving integration compatibility in layered-material devices. 
Compared with topological responses, the thickness-independent responses proposed here may also offer a broader range of candidate materials and greater robustness over temperature and doping.
Beyond device functionality, thickness-dependent responses provide a new route for identifying magnetic structures. Nonlinear optical responses have been widely adopted to probe magnetic symmetries \cite{fiebig2005magneto}; however, different magnetic configurations can sometimes share the same global magnetic symmetry. For instance, A-type AFM with interlayer AFM configuration and C-type AFM with intralayer AFM configuration may belong to the same magnetic symmetry group. According to our analysis, only the interlayer AFM configuration can exhibit thickness-independent, surface-dominated responses. Thus, these two AFM configurations can be distinguished through thickness-dependent or surface-sensitive measurements, as illustrated in Fig.~\ref{Fig4}(c). Moreover, since the thickness-dependent response behavior is governed by the stacking sequence, different stacking orders may also be identified by the thickness-dependent response properties.

\textit{Conclusion.}---In summary, we reveal a general mechanism for realizing thickness-independent quantum geometric responses in layered materials through symmetry breaking induced by interlayer antiferroic coupling.
We propose symmetry requirements and potential material candidates for different kinds of thickness-independent quantum geometric responses. Taking layered AFM MnS as the representative, we show that interlayer AFM coupling-induced $\mathcal{PT}$ symmetry breaking can lead to a surface-dominated AHE, whose magnitude and thickness-independent behavior can be effectively controlled by the stacking order. Our results establish a symmetry-based route for engineering robust quantum geometric responses without relying on band topology, opening new possibilities for device design based on layered antiferroic materials.

\begin{acknowledgments}
This work is supported by the National Key R\&D Program of China (Grant No. 2021YFA1401600), the National Natural Science Foundation of China (Grant No. 12474056).  The computational resources were supported by the high-performance computing platform of Peking University. 
\end{acknowledgments}


\nocite{rauch2018geometric,kresse1996efficient,blochl1994projector,perdew1996generalized,dudarev1998electron,
pizzi2020wannier90,wu2018wanniertools}
\bibliography{ref}
\end{document}